\documentclass[twocolumn,superscriptaddress]{revtex4}
\usepackage{graphicx}
\usepackage{epsfig}
\usepackage{amsmath,amssymb}
\usepackage{amsfonts,amsbsy}
\usepackage{hyperref}
\usepackage{bm}
\usepackage{color}
\usepackage{enumerate}
\usepackage{verbatim}
\usepackage[utf8]{inputenc}
\usepackage{graphicx}
\usepackage{multirow}
\usepackage{lipsum}

\def\be{\begin{equation}}
\def\ee{\end{equation}}
\def\bea{\begin{eqnarray}}
\def\eea{\end{eqnarray}}

\newcommand{\bo}[1]{\boldsymbol{#1}}

\begin{document}
\title{CGC-induced longitudinal ridge in p-Pb collisions}

\author{Donghai Zhang}
\email{zdh@cwnu.edu.cn}
\affiliation{School of Physics and Astronomy, China West Normal Universty, Nanchong 637002, China}
\author{Yeyin Zhao}
\affiliation{School of Physics and Electronic Engineering, Sichuan University of Science and Engineering, Zigong 643000, China}
\author{Luhua Qiu}
\affiliation{Key Laboratory of Quark and Lepton Physics (MOE) and Institute of Particle Physics, Central China Normal University, Wuhan 430079, China}
\author{Mingmei Xu}
\email{xumm@ccnu.edu.cn}
\affiliation{Key Laboratory of Quark and Lepton Physics (MOE) and Institute of Particle Physics, Central China Normal University, Wuhan 430079, China}
\author{Yuanfang Wu}
\email{wuyf@ccnu.edu.cn}
\affiliation{Key Laboratory of Quark and Lepton Physics (MOE) and Institute of Particle Physics, Central China Normal University, Wuhan 430079, China}
\begin{abstract}
Within the Color Glass Condensate (CGC) effective field theory, we investigate the long-range rapidity correlations in proton-lead (p-Pb) collisions at $\sqrt{s_{\mathrm{NN}}}=5.02$ TeV. A distinctive correlation rebound is observed, where the correlation bounces after reaching a minimum at large rapidity gaps ($|\Delta \eta|>2$). The rebound means a strong correlation appears at large rapidity gap. Studying the rebound structures can thus illuminate the formation of the ridge. We find that the rebound is most obvious when the transverse momenta of two measured particles are around 2 $\mathrm{GeV/c}$, and it moves to larger rapidity gaps at higher collision energies. Beyond that, the rapidity correlations in p-Pb collisions show asymmetry when the transverse momenta of two particles differ. The asymmetry, a unique signature of the asymmetric collisions, vanishes when the transverse momenta of two particles coincide. These findings provide direct insight into gluon saturation and quantum evolution.
\end{abstract}

\maketitle
\section{Introduction}
Two-particle correlations are a powerful tool to explore the mechanism of particle production in hadrons and nucleus collisions at high energy. Such studies involve measuring the distributions of relative azimuthal angle $\Delta\phi$ and relative pseudorapidity $\Delta\eta$ between pairs of particles. The $\Delta\eta$-$\Delta\phi$ correlation functions for photonuclear ($\gamma$A) collisions~\cite{ATLAS-2021-rA}, proton-proton (pp)~\cite{CMS-2010-pp,CMS-2011-pp-7,ATLAS-2015-pp-13,CMS-2016-pp-13,ALICE-2017-pp-7,CMS-2017-pp-13}, proton-nucleus (pA)~\cite{CMS-2013-pPb-5.02,ALICE-2013-pPb-5.02,ATLAS-2013-pPb-5.02,ATLAS-2013-pPb-5.02-2,CMS-2013-pPb-PbPb,ALICE-2016-pPb-5.02} and nucleus-nucleus (AA) collisions~\cite{CMS-2011-pp-PbPb,CMS-2011-PbPb-2.76,CMS-2012-PbPb-2.76,PHENIX-2008,STAR-2009,PHOBOS-2010}, show similar structures. An enhancement is observed on the near side (relative azimuthal angle $\Delta\phi\approx0$), which extends over a broad range in relative pseudorapidity ($|\Delta\eta|\approx4$). Such a long-range correlation of the near side is often referred to as the ``ridge". The pp collision data~\cite{CMS-2011-pp-7} and our previous work~\cite{Zhang-2} indicate that the ridge has fine structures in rapidity direction, i.e., the correlation bounces after reaching a minimum at $|\Delta \eta|>2$, which is called a ``rebound". The rebound structure means a strong correlation appears at large rapidity gap. Further insight into the production mechanism of ridge structures may be gained by studying the energy and transverse momentum dependence of rebound structures.

The origin of ridge correlations in high-energy small system collisions has sparked intense debate between two competing mechanisms. One is the CGC effective field theory based on initial-state intrinsic momentum anisotropy~\cite{Dumitru-2008,Dusling-2010,Dumitru-2011,Dusling-2012,Dusling-2013-1,Dusling-2013-2,Dusling-2013-3,Mace-2019,CGC-gammaA}, and the other is final-state interaction of collective response to the collision geometry described by hydrodynamics~\cite{Werner-2011,Bozek-2013,Bzdak-2013,Qin-2014,Werner-2014}. 

The hydrodynamic model, using fluid evolution and various initial conditions, gives a systematical description of the ridge phenomenon in small systems and heavy-ion collisions~\cite{Weller-2017,Nagle-2018}. Several other observations, like mass-ordering of the elliptic flow coefficient ($v_2$) of identified particles~\cite{Alice-PLB2013}, quark-number scaling of $v_2$~\cite{CMS-PLB742} and baryon-to-meson enhancement at intermediate $p_{\rm T}$~\cite{Alice-PLB728} suggest similarity in small systems and heavy-ion collisions. However, the model fails to reproduce multiple-particle cumulant $c_2\{4\}$~\cite{hydro-c24}. Moreover, the absence of significant jet-quenching indicates that QGP medium may not exist in small systems~\cite{Alice-PRL2013,Alice-PLB2016}, in which case hydrodynamics may not be applicable.

On the other hand, CGC successfully explained the mass ordering of $v_2$~\cite{CGC-mass}, multiple-particle cumulant $c_2\{4\}$~\cite{CGC-c24}, and the elliptic flow of $\gamma$A process~\cite{CGC-gammaA}, except the ordering of Fourier harmonics $v_n$ in the three system sizes of p-Au, d-Au and He-Au~\cite{CGC-fail}. Furthermore, sizable signals of collectivity have been found not only for soft and light hadrons but also for heavy flavor mesons~\cite{Alice-PLB2018,CMS-flavor,CMS-PRL2018,CMS-19} in small systems. In this respect, CGC successfully explained experimental data~\cite{CGC-flavor}, hydrodynamics failed~\cite{hydro-flavor}. In brief, either hydrodynamics or CGC might not rule data all. Recent research~\cite{latest-1,latest-2,latest-3} shows that the azimuthal anisotropy measured in small systems may result from the coexistence of CGC and hydrodynamics.

The gluon density inside hadron or nucleus sharply grows at small Bjorken $x$ or high energy. The framework to describe the physics of high parton densities and gluon saturation inside projectile and target hadron or nucleus is the CGC effective field theory~\cite{CGC-review1,CGC-review2,CGC-review3,CGC-review}. The effective degrees of freedom in this framework are color sources $\rho$ at large $x$ and gauge fields $\mathcal{A}_{\mu}$ at small $x$. The classical gauge field $\mathcal{A}_\mu$ is the solution of classical Yang-Mills equations with a fixed configuration of color sources. For a given initial configuration of color source, field in the nuclear wave functions evolve with Bjorken $x$, which is described by the Jalilian-Marian-Iancu-McLerran-Weigert-Kovner (JIMWLK) renormalization group equations~\cite{rcBK-1,JIMWLK-1,JIMWLK-2,JIMWLK-22,JIMWLK-3}. In the mean field approximation and large-$N_c$ limit, the JIMWLK equation is reduced to the Balitsky-Kovchegov (BK) equation~\cite{rcBK-1,rcBK-2,rcBK-3}. 

The projectile and target hadron or nucleus of high gluon density are two sheets of strongly correlated coherent gluonic fields, which is called color glass condensate. When two sheets of CGCs shatter in a high-multiplicity collision, the strong longitudinal color electric and color magnetic fields are formed in the colliding region and are called glasma~\cite{glasma}. Particles with large rapidity separation are produced locally in the transverse plane and correlated by approximately boost invariant glasma flux tubes, generated the long-range rapidity correlations.  

In high energy limit, the Bjorken $x$ of a gluon is related to the transverse momentum $p_{\rm T}$, center-of-mass energy $\sqrt{s}$ and rapidity $y$,
\begin{equation}
x=\frac{p_{\rm T}}{\sqrt{s}}{\rm e}^{\pm y},
\end{equation} 
where $+(-)$ corresponds to the right(left) moving projectile(target). The gluon correlations are governed by unintegrated gluon distribution (uGD) which peaks around the saturation momentum $Q_{\rm s}$ (see Fig~.1 in ref.~\cite{Zhao-1}). Within the CGC framework, $Q_{\rm s}$ is determined by $x$ for a given colliding hadron (Fig.~2(a) in ref.~\cite{Zhao-1}), the uGD consequently depends on $x$. Physically, gluons with different $x$ degree of freedom reflect different stages of partonic evolution in a colliding hadron and correspond to different rapidity regions (Fig.~1(b) in ref.~\cite{Zhang-1}). Conversely, gluons at different rapidities effectively represent different “generations” in the quantum evolution. Large-rapidity ridge correlations thus probe strong correlations between  large $x$  source gluons and small $x$ radiated descendants. Eq.~(1) shows transverse momentum and collision energy $\sqrt{s}$ have a direct influence on Bjorken $x$. Large-rapidity ridge correlations are thus sensitive to the collision energy and transverse momenta of correlated gluons.
 
In the CGC calculation, we correct the number of uncorrelated pairs as the integral of the product of two single-particle distributions within the acceptance~\cite{Zhang-2}. Using this correction in p-Pb collisions, the ridge yield at large rapidity rebounds after bottoming. This agrees with the trend in pp data~\cite{CMS-2011-pp-7}, based on the current lack of p-Pb data and the similar ridge structure in pp and p-Pb collisions. A key objective of this paper aims to provide theoretical predictions and call for new experimental measurements. The correlation rebound is found most obvious around the sum of the saturation momenta of the projectile and target, and moves to the larger rapidity gap for higher colliding energies. The strongest gluon correlation occurs at the peak of the uGD which depends on $x$. At fixed $x$, Eq.~(1) shows that the rapidity $y_q$ increases while $y_p$ decreases as $\sqrt{s}$ increases, the dependence on rapidity is much more sensitive due to the exponential function. Physically, this means the observed correlation rebound is a direct signature of strong correlations between large $x$ source gluons and their small $x$ radiated descendants. 

It is further observed that long-range rapidity correlations show a strong asymmetry in p-Pb systems when the transverse momenta of two particles differ, but become symmetric when the transverse momenta coincide. In contrast, such rapidity correlations in pp collisions remain symmetric regardless of whether the transverse momenta differ or not. The asymmetry in rapidity correlations stems from the inherent asymmetry of the colliding hadron or nucleus in rapidity. The long-range rapidity correlation is determined by the uGD which depends on $x$. According to Eq.~(1), $x$ is exclusively a function of the transverse momentum $p_{\rm T}$ for fixed $y$ and $\sqrt{s}$. Therefore, the asymmetry in rapidity correlations are sensitive to the transverse momenta of correlated gluons.

This paper is organized as follows. The definition of correlation function and some related formulae in CGC framework are given in section II. The formulae in this manuscript follow those in refs.~\cite{Dusling-2013-1,Dusling-2013-2,Dusling-2013-3,Zhao-1,ZhangHY,Zhang-1,Zhao-2,Zhang-2} and are identical with those at gluon level without fragmentation functions. In section III, the per-trigger yield in the $\Delta y$-$\Delta\phi$ plane for 5.02 TeV p-Pb collisions is presented. In section IV, the dependence of large-rapidity correlations on transverse momentum is systematically studied and the origin is discussed. Section V is a brief summary and discussions.  

\section{Two-particle correlations in the framework of CGC}

The per-trigger yield is defined as
\begin{equation}
Y(\Delta y,\Delta\phi)=\frac{1}{N_{\rm Trig}} \frac{\mathrm{d}^2 N^{\rm pair}}{\mathrm{d}\Delta y\mathrm{d}\Delta\phi}=B(0,0)\frac{S(\Delta y,\Delta\phi)}{B(\Delta y,\Delta\phi)}.
\end{equation}
It counts the number of correlated particle pairs with rapidity separation($\Delta y=y_q-y_p$) and azimuthal angle separation($\Delta\phi=\phi_q-\phi_p$), divided by the number of trigger particles. 

 The functions $S(\Delta y,\Delta\phi)$ and $B(\Delta y,\Delta\phi)$ are particle pairs obtained from original and mixed events, respectively. They are known as the signal and background distributions, denoted as
\begin{eqnarray}
S(\Delta y,\Delta\phi)=\frac{1}{N_{\rm Trig}} \frac{\mathrm{d}^2 N^{\rm same}}{\mathrm{d}\Delta y\mathrm{d}\Delta\phi},\\
B(\Delta y,\Delta\phi)=\frac{1}{N_{\rm Trig}} \frac{\mathrm{d}^2 N^{\rm mixed}}{\mathrm{d}\Delta y\mathrm{d}\Delta\phi},
\end{eqnarray}  
respectively. 

The signal distributions $S(\Delta y,\Delta\phi)$ in Eq. (3), contain not only correlated pairs but also uncorrelated pairs.

In experiments, the number of uncorrelated pairs is estimated by mixed events. Particles of a mixed event are drawn randomly from different original events. For a large enough number of original events, in a single mixed event, the probability of having two particles from the same original event is close to zero. The particles in a mixed event are almost independent~\cite{Gazd}.  

Detector effects, such as tracking inefficiency and detector acceptance, can also affect the signal distributions. They largely cancel in the $S(\Delta y,\Delta\phi)/B(\Delta y,\Delta\phi)$ ratio. The factor $B(0,0)$ is the value of $B(\Delta y,\Delta\phi)$ at $\Delta y=0$ and $\Delta\phi=0$. 

Experiment samples individual events, while theory provides the properties of statistical distributions. In CGC theory, the signal distribution is obtained by the integration of double-particle inclusive distributions. It is expressed as~\cite{Dusling-2013-3,Dusling-2013-1,Dusling-2013-2} 
\begin{eqnarray}
S^{'}(\Delta y,\Delta\phi)=\frac{1}{N_{\rm Trig}} \frac{\mathrm{d}^2 N^{\rm assoc}}{\mathrm{d}\Delta y\mathrm{d}\Delta\phi},
\end{eqnarray}
where
\begin{widetext}
\begin{eqnarray}
\frac{\mathrm{d}^{2}N^{\rm{assoc}}}{\mathrm{d}\Delta y \mathrm{d}\Delta\phi} &=&\int_{y^{\rm min}-y_{\rm shift}}^{y^{\rm max}-y_{\rm shift}} \mathrm{d}y_{\rm p}\int_{y^{\rm min}-y_{\rm shift}}^{y^{\rm max}-y_{\rm shift}} \mathrm{d}y_{\rm q} \delta(y_{\rm q}-y_{\rm p}-\Delta y)\int_0^{2\pi} \mathrm{d}\phi_{\rm p} \int_0^{2\pi} \mathrm{d}\phi_{\rm q}  \delta(\phi_{\rm q}-\phi_{\rm p}-\Delta \phi) \nonumber\\ &&\times\int^{p^{\rm{max}}_{\rm T}}_{ p^{\rm{min}}_{\rm T}}\frac{\mathrm{d}p^{2}_{\rm T}}{2}\int^{q^{\rm{max}}_{\rm T}}_{q^{\rm{min}}_{\rm T}}\frac{\mathrm{d}q^{2}_{\rm T}}{2}\frac{\mathrm{d}N^{\rm{corr}}_{\mathrm{2}}}{{\mathrm{d}^{2}\boldsymbol{p}_{\rm T} \mathrm{d}y_{\rm p}}{\mathrm{d}^{2}\boldsymbol{q}_{\rm T} \mathrm{d}y_{\rm q}}}.
\end{eqnarray}\end{widetext}
The labels ``p" and ``q" denote the two particles in the pair, corresponding to ``trigger" and ``associated" particles in experiment, respectively. $y_{\rm shift}=0.465$ is the shift in rapidity in the center-of-mass frame towards the lead fragmentation region in asymmetric p-Pb collisions. The $\delta$ function is used to restrict the phase space interval to a given $\Delta y$ and $\Delta\phi$. The correlated double-particle inclusive distributions $\frac{\mathrm{d}N^{\rm{corr}}_{\mathrm{2}}}{{\mathrm{d}^{2}\boldsymbol{p}_{\rm T} \mathrm{d}y_{\rm p}}{\mathrm{d}^{2}\boldsymbol{q}_{\rm T} \mathrm{d}y_{\rm q}}}$ is equal to the double-particle inclusive distributions minus the product of two single-particle inclusive distributions, i.e.,   
\begin{widetext}
\begin{equation}
\frac{\mathrm{d}N^{\rm{corr}}_{\mathrm{2}}}{{\mathrm{d}^{2}\boldsymbol{p}_{\rm T} \mathrm{d}y_{\rm p}}{\mathrm{d}^{2}\boldsymbol{q}_{\rm T} \mathrm{d}y_{\rm q}}}=\frac{\mathrm{d}N_{\mathrm{2}}}{{\mathrm{d}^{2}\boldsymbol{p}_{\rm T} \mathrm{d}y_{\rm p}}{\mathrm{d}^{2}\boldsymbol{q}_{\rm T} \mathrm{d}y_{\rm q}}}-\frac{\mathrm{d}N_1}{\mathrm{d}^{2}\boldsymbol{p}_{\rm T} \mathrm{d}y_{\rm p}}\frac{\mathrm{d}N_1}{\mathrm{d}^{2}\boldsymbol{q}_{\rm T} \mathrm{d}y_{\rm q}}.
\end{equation} 
\end{widetext}

In the CGC framework, the observable under the leading log approximation is factorized as~\cite{Dumitru-2008}
\begin{equation}\label{eq:llog}
\langle\mathcal{O}\rangle_{\mathrm{LLog}}=\int\left[D\rho_1\right]\left[D\rho_2\right]W[\rho_1]W[\rho_2]\mathcal{O}[\rho_1,\rho_2]_{\mathrm{LO}},
\end{equation}
where $\mathcal{O}[\rho_1,\rho_2]_{\mathrm{LO}}$ is the leading order single- or double- particle inclusive distribution for a fixed distribution of color sources. The integration denotes an average over different distributions of the color sources with the weight functional $W[\rho_{1,2}]$. In general, $W[\rho_{1,2}]$ encodes all possible color charge configurations of projectile and target, and obeys JIMWLK renormalization group equations\cite{JIMWLK-1,JIMWLK-2,JIMWLK-3}. Where all quantum information of projectile/target is absorbed into the distribution $W[\rho_{1,2}]$.

The averaging over color sources can be done under MV model with a Gaussian weight functional. According to refs.~\cite{Dusling-2010,Zhao-2}, the two-gluon inclusive distribution can be expressed as
\begin{widetext}
\begin{eqnarray}\label{eq:22}
\left\langle\frac{dN_2}{d^2\bo{p}_{\rm T} dy_p d^2\bo{q}_{\rm T} dy_q}\right\rangle&=&\frac{\alpha^2_s N^2_c S_\perp}{\pi^{8}(N^2_c-1)^3}\frac{1}{\bo{p}^2_{\rm T}\bo{q}^2_{\rm T}}\int \frac{d^2\bo{k}_{\rm T}}{(2\pi)^2} \Bigg[(D_1+D_2)+\sum_{j=\pm}\left[D_3(\bo{p}_{\rm T},j\bo{q}_{\rm T})+\frac{1}{2}D_4(\bo{p}_{\rm T},j\bo{q}_{\rm T}\right]\Bigg]
\nonumber\\ &&+\left\langle\frac{dN_1}{d^2\bo{p}_{\rm T} dy_p}\right\rangle \left\langle\frac{dN_1}{d^2\bo{q}_{\rm T} dy_q}\right\rangle,
\end{eqnarray}
\end{widetext}
where
\begin{eqnarray}\label{terms}
D_1= \Phi^2_{A}(y_p,\bo{k}_{\rm T})\Phi_{B}(y_p,\bo{p}_{\rm T}-\bo{k}_{\rm T})D_B,\nonumber\\
D_2= \Phi^2_{B}(y_q,\bo{k}_{\rm T})\Phi_{A}(y_p,\bo{p}_{\rm T}-\bo{k}_{\rm T})D_A,
\end{eqnarray}
with
\begin{equation}\label{eq:5}
D_{A(B)}=\Phi_{A(B)}(y_q,\bo{q}_{\rm T}+\bo{k}_{\rm T})
+\Phi_{A(B)}(y_q,\bo{q}_{\rm T}-\bo{k}_{\rm T}).
\end{equation}
Here $\Phi_{A(B)}(y,\bo{k}_{\rm T})$ denotes uGD of projectile $A$ or target $B$, $S_\perp$ is the transverse overlap area of two uGDs. Where $D_3(\bo{p}_{\rm T},j\bo{q}_{\rm T})=\delta^2(\bo{p}_{\rm T} - j\bo{q}_{\rm T})\left[\mathcal{I}^2_1 + \mathcal{I}^2_2 + 2\mathcal{I}^2_3\right]$, with
\begin{widetext}
\begin{eqnarray}\label{eq:d3}
\mathcal{I}_1= \int\frac{d^2\bo{k}_{\rm T}}{(2\pi)^2}
\Phi_{A_1}(y_p,\bo{k}_{\rm 1T})\Phi_{A_2}(y_q,\bo{p}_{\rm T}-\bo{k}_{\rm 1T})
\frac{(\bo{k}_{\rm 1T}\cdot\bo{p}_{\rm T} - \bo{k}^2_{\rm 1T})^2}{\bo{k}^2_{\rm 1T}(\bo{p}_{\rm T}-\bo{k}_{\rm 1T})^2},\nonumber\\
\mathcal{I}_2=\int\frac{d^2\bo{k}_{\rm 1T}}{(2\pi)^2}
\Phi_{A_1}(y_p,\bo{k}_{\rm 1T})\Phi_{A_2}(y_q,\bo{p}_{\rm T}-\bo{k}_{\rm 1T})
\frac{\vert\bo{k}_{\rm 1T}\times\bo{p}_{\rm T}\vert^2}{\bo{k}^2_{\rm 1T}(\bo{p}_{\rm T}-\bo{k}_{\rm 1T})^2},\nonumber\\
\mathcal{I}_3=\int\frac{d^2\bo{k}_{\rm 1T}}{(2\pi)^2}
\Phi_{A_1}(y_p,\bo{k}_{\rm 1T})\Phi_{A_2}(y_q,\bo{p}_{\rm T}-\bo{k}_{\rm 1T})
\frac{(\bo{k}_{\rm 1T}\cdot\bo{p}_{\rm T})\vert\bo{k}_{\rm 1T}\times\bo{p}_{\rm T}\vert}{\bo{k}^2_{\rm 1T}(\bo{p}_{\rm T}-\bo{k}_{\rm 1T})^2},\nonumber\\
\end{eqnarray} 
\end{widetext}
and
\begin{widetext}
\begin{eqnarray}\label{eq:d4}
D_4(\bo{p}_{\rm T},j\bo{q}_{\rm T})&=&\int\frac{d^2\bo{k}_{\rm 1T}}{(2\pi)^2}
\Phi_{A_1}(y_p,\bo{k}_{\rm 1T})
\Phi_{A_1}(y_p,\bo{k}_{\rm 2T})
\Phi_{A_2}(y_q,\bo{p}_{\rm T}-\bo{k}_{\rm 1T})
\Phi_{A_2}(y_q,\bo{p}_{\rm T}-\bo{k}_{\rm 2T})\nonumber\\ &&\times\frac{(\bo{k}_{\rm 1T}\cdot\bo{p}_{\rm T}-\bo{k}^2_{\rm 1T})
(\bo{k}_{\rm 2T}\cdot\bo{p}_{\rm T}-\bo{k}^2_{\rm 2T})
+(\bo{k}_{\rm 1T}\times\bo{p}_{\rm T})\cdot(\bo{k}_{\rm 2T}\times\bo{p}_{\rm T})}{\bo{k}^2_{\rm 1T}(\bo{p}_{\rm T}-\bo{k}_{\rm 1T})^2}
\nonumber\\ &&\times
\frac{(\bo{k}_{\rm 1T}\cdot j\bo{q}_{\rm T}-\bo{k}^2_{\rm 1T})
(\bo{k}_{\rm 2T}\cdot j\bo{q}_{\rm T}-\bo{k}^2_{\rm 2T})
+(\bo{k}_{\rm 1T}\times\bo{q}_{\rm T})\cdot
(\bo{k}_{\rm 2T}\times\bo{q}_{\rm T})}
{\bo{k}^2_{\rm 2T}(j\bo{q}_{\rm T}-\bo{k}_{\rm 1T})^2},
\end{eqnarray}
\end{widetext}
where $\bo{k}_{\rm 2T}=\bo{p}_{\rm T}+j\bo{q}_{\rm T}-\bo{k}_{\rm 1T}$, $j=\pm$. 

By utilizing this factorization formalism (i.e. Eq.~(8)), the single-gluon inclusive distribution is directly written as,
\begin{flalign}\label{eq:11}
&\left\langle\frac{dN_1}{d^2\bo{p}_{\rm T} dy_p}\right\rangle\nonumber\\ =&\frac{\alpha_s N_c S_\perp}{\pi^4(N^2_c-1)}\frac{1}{\bo{p}^2_{\rm T}}\int\frac{d^2\bo{k}_{\rm T}}{(2\pi)^2}\Phi_{A}(y_p,\bo{k}_{\rm T})\Phi_{B}(y_p,\bo{p}_{\rm T}- \bo{k}_{\rm T}).
\end{flalign} 
The number of trigger particles reads
\begin{eqnarray}
N_{\rm{Trig}} &=& \iiint_{\mathrm{Acceptance}}dy d^{2}\mathbf{p}_{\rm T}\frac{dN_1}{d^{2}\mathbf{p}_{\rm T} dy_{\rm p}}.
\end{eqnarray}
The framework is valid to leading logarithmic accuracy in $x$ and momenta $p_{\rm T}, q_{\rm T} \gg Q_{\rm s}$. Eq.~(9), (14) are well-known $k_{\rm T}$-factorization formalism~\cite{kT} which is widely used to compute the double- or single-particle inclusive distributions at high energy collisions. The convolution of uGDs ($\Phi$) in the single- and double-particle inclusive distributions reflects the initial correlations between gluons from the two colliding nucleus. The correlations originate from the statistical properties and quantum evolution of the color sources.

The background distributions in Eq. (4) represent the distribution of uncorrelated pairs. The counterpart in theoretical calculations should be integrals of the product of two single-particle inclusive distributions, i.e.,  
\begin{eqnarray}
B^{'}(\Delta y,\Delta\phi)=\frac{1}{N_{\rm Trig}} \frac{\mathrm{d}^2 N^{\rm uncorr}}{\mathrm{d}\Delta y\mathrm{d}\Delta\phi},
\end{eqnarray}
with
\begin{widetext}
\begin{eqnarray}
\frac{\mathrm{d}^{2}N^{\rm{uncorr}}}{\mathrm{d}\Delta y \mathrm{d}\Delta\phi} &=&\int_{y^{\rm min}-y_{\rm shift}}^{y^{\rm max}-y_{\rm shift}} \mathrm{d}y_{\rm p}\int_{y^{\rm min}-y_{\rm shift}}^{y^{\rm max}-y_{\rm shift}} \mathrm{d}y_{\rm q}\delta(y_{\rm q}-y_{\rm p}-\Delta y)\int_0^{2\pi} \mathrm{d}\phi_{\rm p} \int_0^{2\pi} \mathrm{d}\phi_{\rm q}  \delta(\phi_{\rm q}-\phi_{\rm p}-\Delta \phi) \nonumber\\ &&\times\int^{p^{\rm{max}}_{\rm T}}_{p^{\rm{min}}_{\rm T}}\frac{\mathrm{d}p^{2}_{\rm T}}{2}\int^{q^{\rm{max}}_{\rm T}}_{q^{\rm{min}}_{\rm T}}\frac{\mathrm{d}q^{2}_{\rm T}}{2}\frac{\mathrm{d}N_1}{\mathrm{d}^{2}\boldsymbol{p}_{\rm T} \mathrm{d}y_{\rm p}}\frac{\mathrm{d}N_1}{\mathrm{d}^{2}\boldsymbol{q}_{\rm T} \mathrm{d}y_{\rm q}}.
\end{eqnarray}\end{widetext}
The integration in Eq.~(17) depends on the shape of the single-particle distribution and the acceptance. The background distribution is the function of $\Delta y$ due to the limited rapidity acceptance. To precisely reproduce the ridge structure observed in experiments, we corrected the background distribution by integrating the product of two real single-particle distributions~\cite{Zhang-2}. Calculating the ratio of $B^{'}(0,0)*S^{'}(\Delta y,\Delta\phi)/B^{'}(\Delta y,\Delta\phi)$, the per-trigger yield is obtained in the CGC framework. 

The uGD ($\Phi$) is related to quark-antiquark dipole forward scattering amplitude $\mathcal{N}$. In large $N_c$ limit~\cite{Dusling-2010},
\begin{widetext}
\begin{eqnarray}
\Phi(x, \mathbf{k}_{\rm T}) = \frac{N_ck^2_{\rm T}}{4\alpha_s}\int{d^2\mathbf{r}_{\bot}e^{i\mathbf{k}_{\rm T}\cdot\mathbf{r}_{\bot}}\left[1-\mathcal{N}_{\text{ad.}}(\mathbf{r}_{\bot}, Y)\right]} = \frac{\pi N_ck^2_{\rm T}}{2\alpha_s}\int dr_{\bot}r_{\bot}J_0(k_{\rm T}r_{\bot})\left[1-\mathcal{N}(r_{\bot},Y)\right]^2,
\end{eqnarray}
\end{widetext}
where $J_0$ is Bessel function, $Y=\ln\frac{x_0}{x}=\ln x_0-\ln x$ is the rapidity interval of produced gluon, $x_0=0.01$ is starting value of small-$x$ dynamic evolution. The amplitude $\mathcal{N}$ is obtained by solving the leading order running coupling Balitsky-Kovchegov (rcBK) equation at a given initial condition~\cite{rcBK-1,rcBK-2,rcBK-3}. Then the double- and single-particle inclusive distributions are available.

In large-$N_c$ limit, the leading order rcBK equation reads
\begin{widetext}
\begin{eqnarray} 
\frac{\partial \mathcal{N}(\mathbf{r}_\bot,x)}{\partial Y} &=& \int d^2\mathbf{r}_{\bot1}K^{\text{run}}(\mathbf{r}_\bot,\mathbf{r}_{\bot1},\mathbf{r}_{\bot2}) \Big[\mathcal{N}(\mathbf{r}_{\bot1},x) + \mathcal{N}(\mathbf{r}_{\bot2},x)-\mathcal{N}(\mathbf{r}_{\bot},x)-\mathcal{N}(\mathbf{r}_{\bot1},x)\mathcal{N}(\mathbf{r}_{\bot2},x)\Big],
\end{eqnarray}\end{widetext}
where $\mathbf{r}_{\bot}={\mathbf{r}_{\bot}}_1+{\mathbf{r}_{\bot}}_2$ is the transverse dipole size, $K^{\text{run}}$ is running coupling kernel. In order to solve the rcBK equation, the Albacete-Armesto-Milhano-Quiroqa-Salgodo (AAMQS) initial condition and a running coupling kernel, Balitsky's prescription, are employed~\cite{Balitsky-1,Balitsky-2}. They are denoted as 
\begin{equation}\label{eq:aamqs}
\mathcal{N}_{\text{AAMQS}}(r,x_0)=1-\exp\Big[-\frac{1}{4}\big(r^2Q^2_s(x_0)\big)^{\gamma}\ln\bigg(e+\frac{1}{r\Lambda}\bigg)\Big],
\end{equation}
\begin{eqnarray} 
K^{\text{run}}(\mathbf{r}_{\bot},{\mathbf{r}_{\bot}}_1,{\mathbf{r}_{\bot}}_2) &=& \frac{N_c\alpha_s({\mathbf{r}_{\bot}}^2)}{2\pi^2}\Bigg[\frac{1}{{\mathbf{r}_{\bot}}_1^2}\Bigg(\frac{\alpha_s({\mathbf{r}_{\bot}}_1^2)}{\alpha_s({\mathbf{r}_{\bot}}_2^2)}-1\Bigg) \nonumber \\ &+&\frac{{\mathbf{r}_{\bot}}^2}{{\mathbf{r}_{\bot}}_1^2 {\mathbf{r}_{\bot}}_2^2} + \frac{1}{{\mathbf{r}_{\bot}}_2^2}\Bigg(\frac{\alpha_s({\mathbf{r}_{\bot}}_2^2)}{\alpha_s({\mathbf{r}_{\bot}}_1^2)}-1\Bigg)\Bigg].\nonumber\\
\end{eqnarray}
The best fit of data requires that the infrared scale $\Lambda=0.241\ \mathrm{GeV}$, the anomalous dimension $\gamma=1.119$ and the initial saturation scale $Q^2_{\rm{sp}}(x_0)=0.168\ \mathrm{GeV}^2$ for proton, $Q^2_{\rm{sA}}(x_0)=0.504\ \mathrm{GeV}^2$ for lead nucleus~\cite{Balitsky-2}. These parameters have successfully described proton structure function $F_2$ vs. $x$~\cite{Balitsky-2}, single-inclusive $p_{\rm T}$ spectra~\cite{rcBK-5} and other experimental data~\cite{Dusling-2012,Dusling-2013-1,Dusling-2013-2,Dusling-2013-3}. 

\begin{figure*}[ht]
\begin{center}
\includegraphics[scale=0.47]{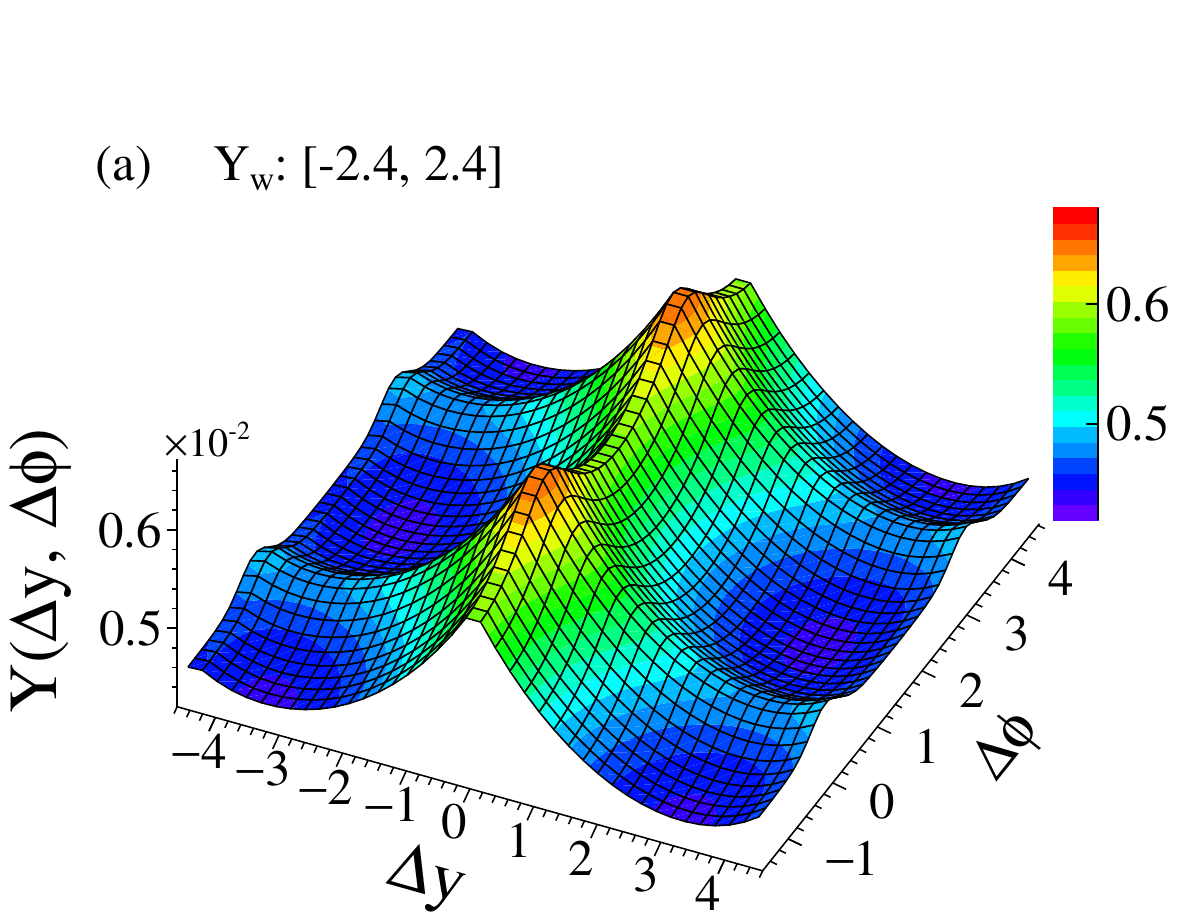}
\includegraphics[scale=0.42]{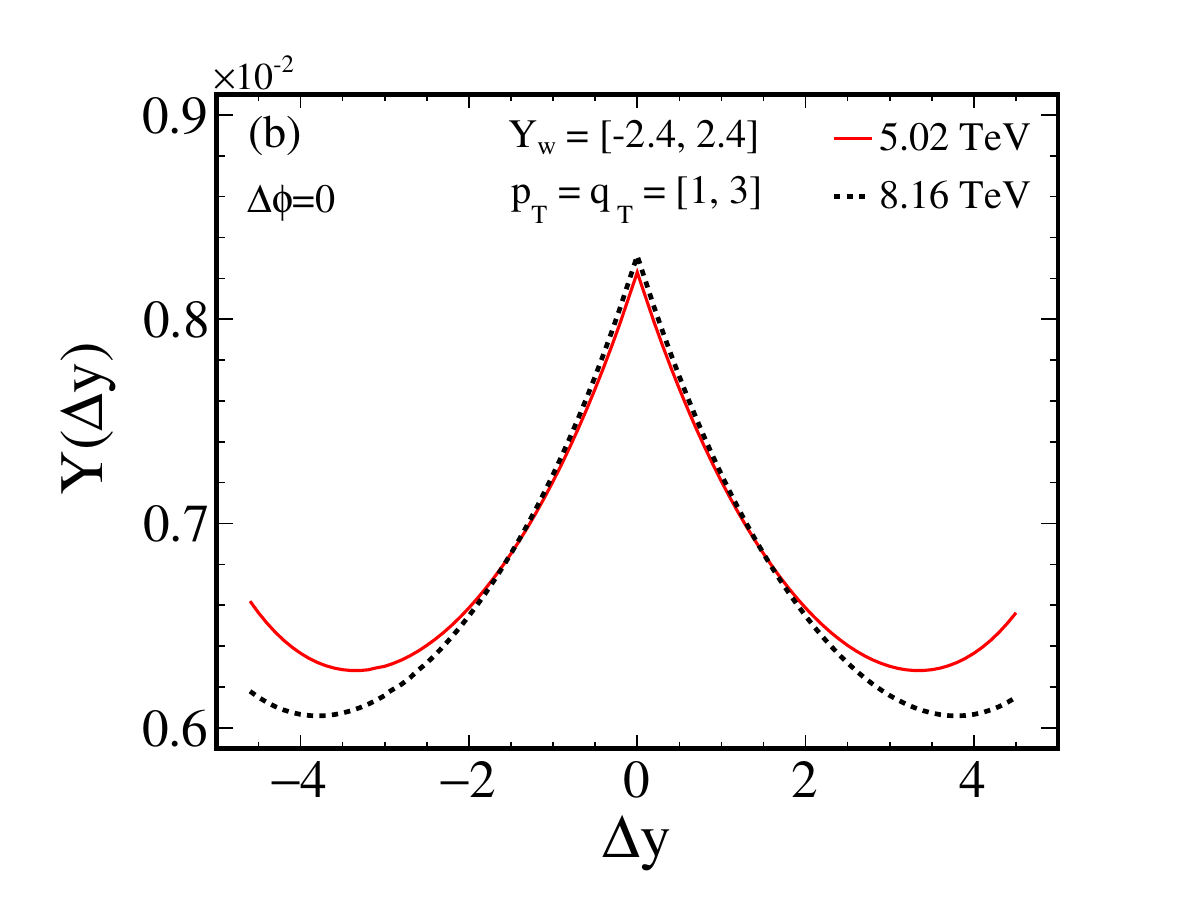}
\end{center}
\caption{The per-trigger yield in the $\Delta y$-$\Delta\phi$ plane for 5.02 TeV p-Pb collisions with transverse momentum integrated within $1\leq p_{\rm T}(q_{\rm T})\leq 3$~GeV$/c$ and with rapidity integrated in $-2.865\leq y_{\rm p}(y_{\rm q})\leq1.935$ (the left panel). The right panel shows the near-side ridge yield projects to $\Delta y$ axis for p-Pb collisions at 5.02 and 8.16 TeV, respectively. }
\end{figure*}
Completing the integrals in Eqs.~(6) and (17) with the transverse momentum range $1\leq p_{\rm T}(q_{\rm T})\leq 3~$GeV$/c$ and the rapidity range $-2.865\leq y_{\rm p}(y_{\rm q})\leq1.935$, $Y(\Delta y,\Delta\phi)$ is obtained and shown in Fig.~1(a). In this Letter, $Y_{\rm w}$ in all figures denotes the rapidity window in the laboratory system.

\section{Long-range rapidity correlations}

\begin{figure*}[htbp]
\begin{centering}
$\vcenter{\hbox{\includegraphics[scale=0.295]{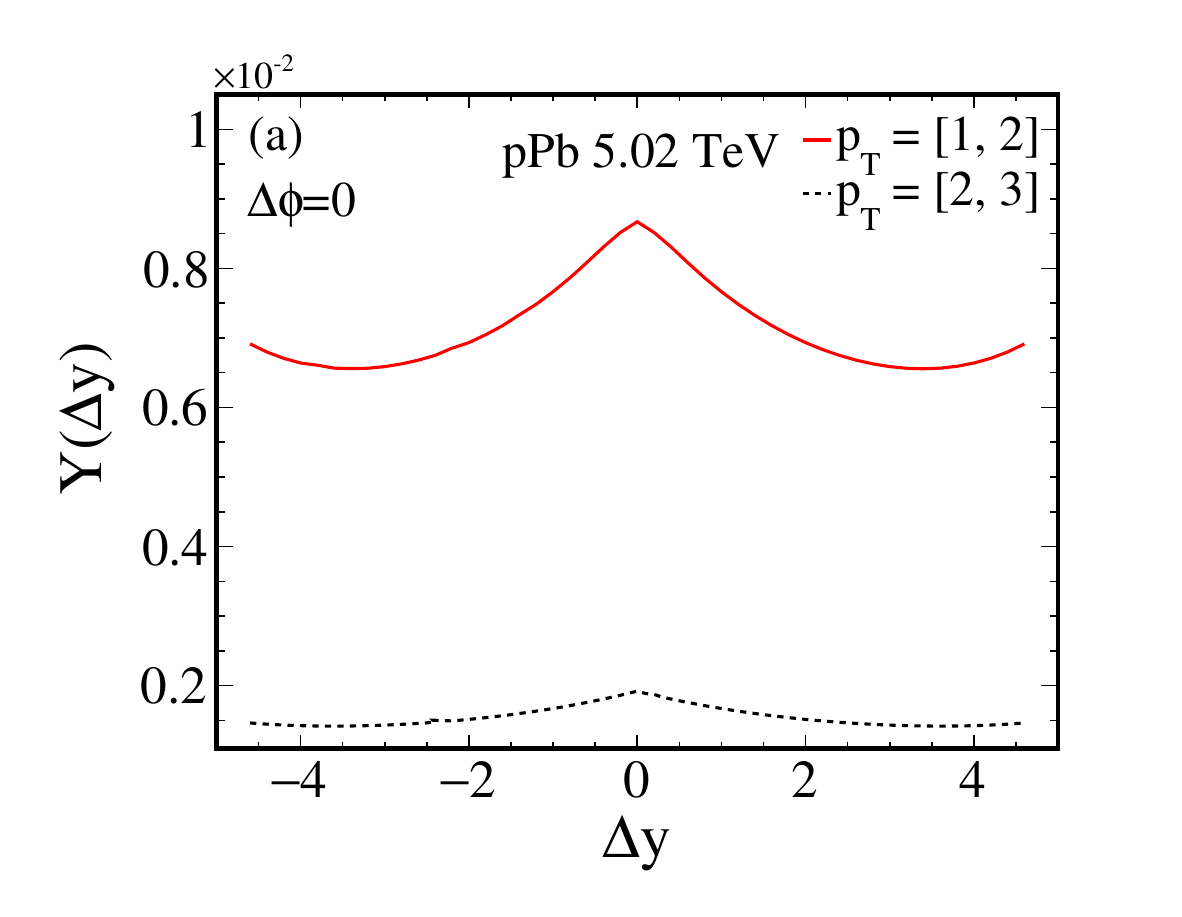}}}$
$\vcenter{\hbox{\includegraphics[scale=0.295]{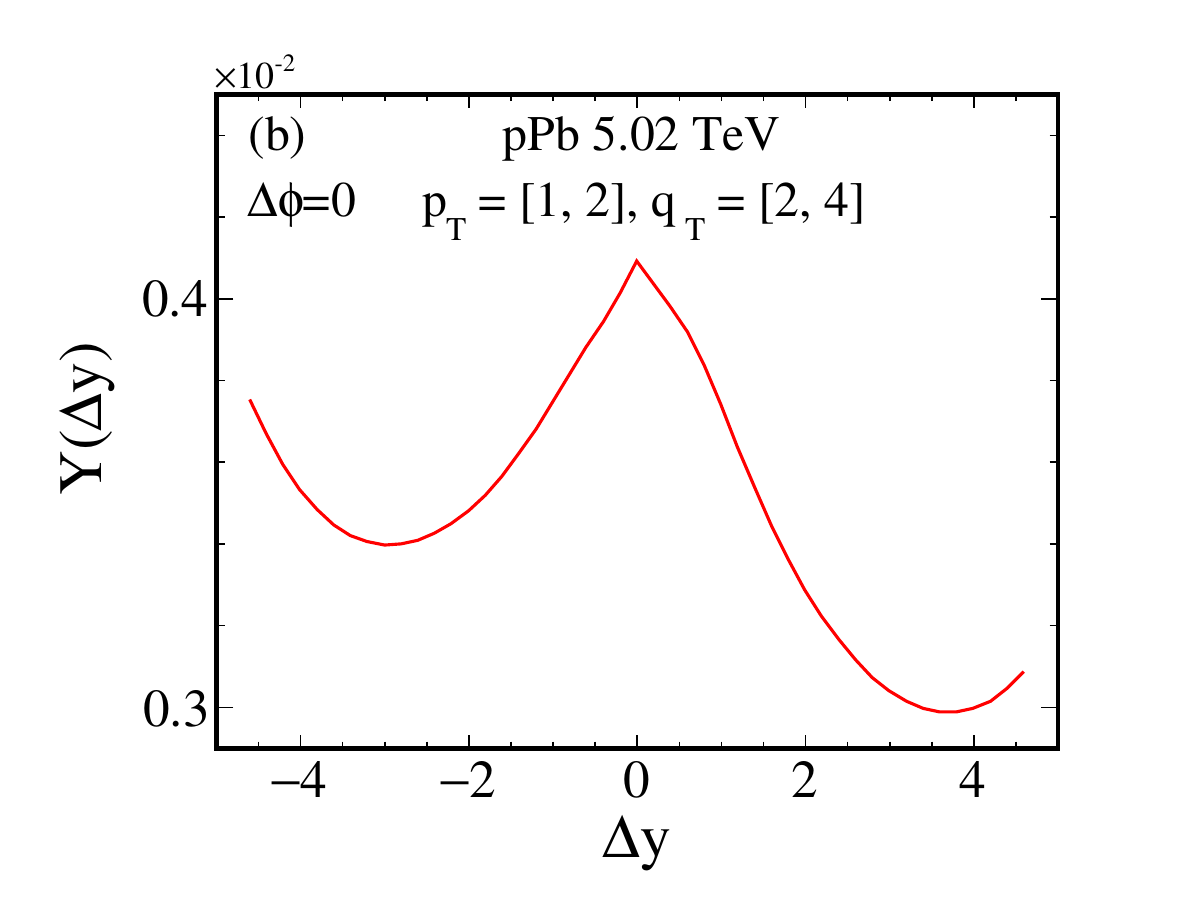}}}$ 
$\vcenter{\hbox{\includegraphics[scale=0.295]{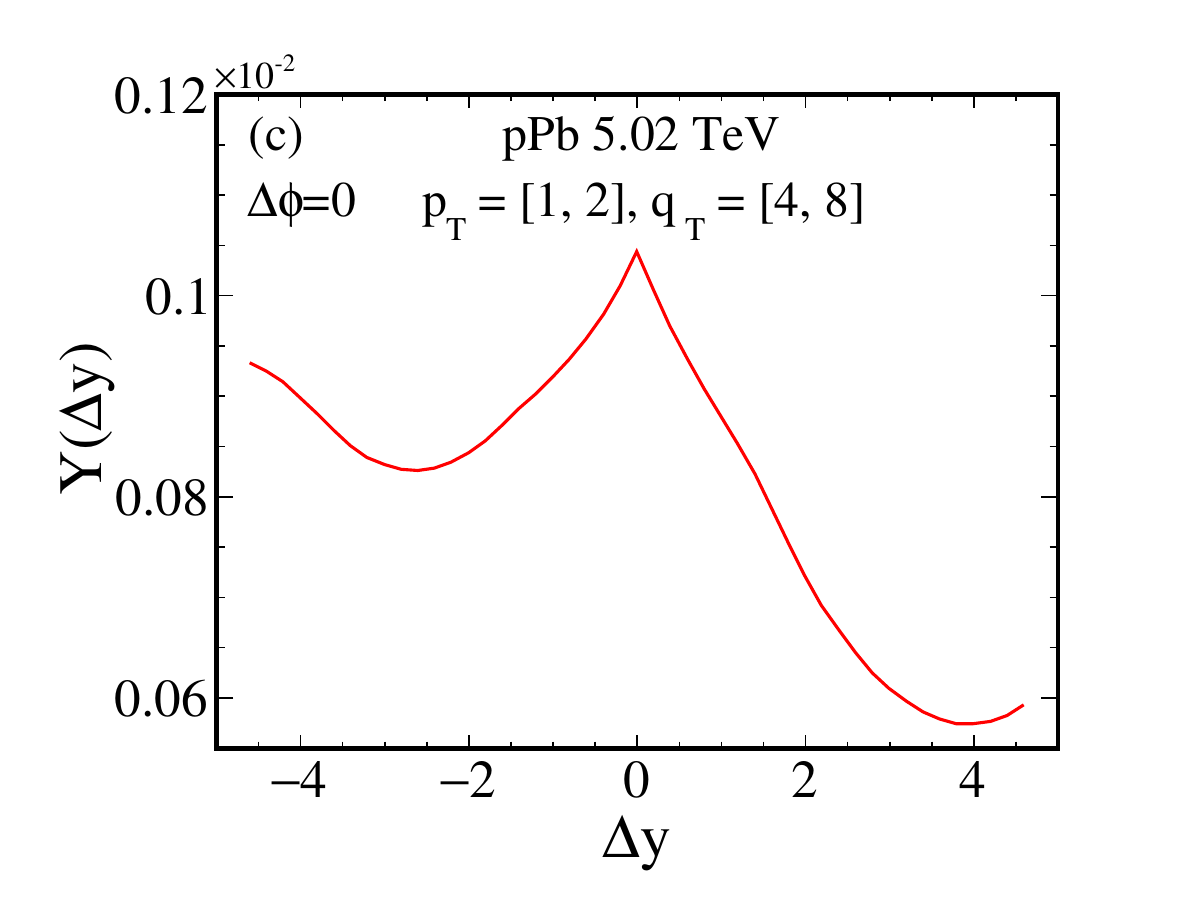}}}$ 
$\vcenter{\hbox{\includegraphics[scale=0.295]{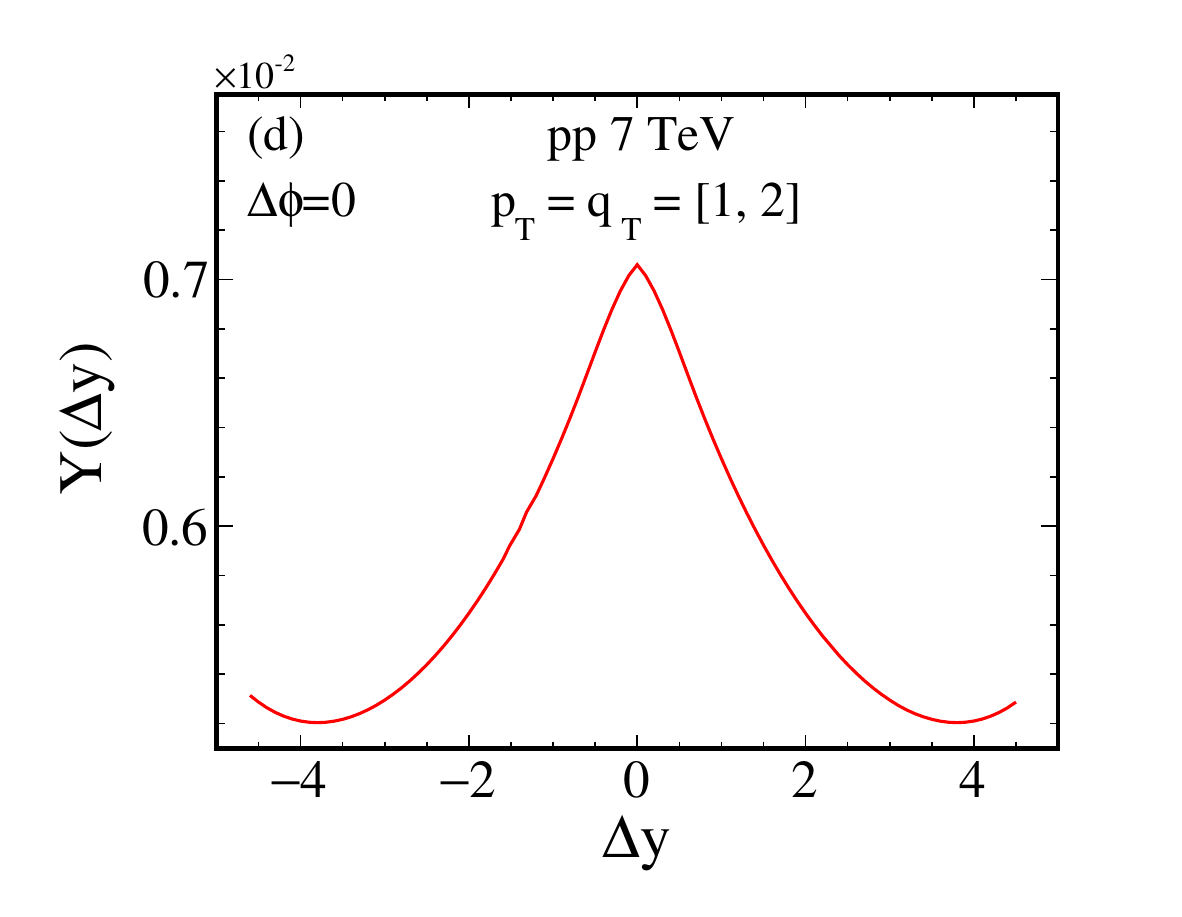}}}$
$\vcenter{\hbox{\includegraphics[scale=0.295]{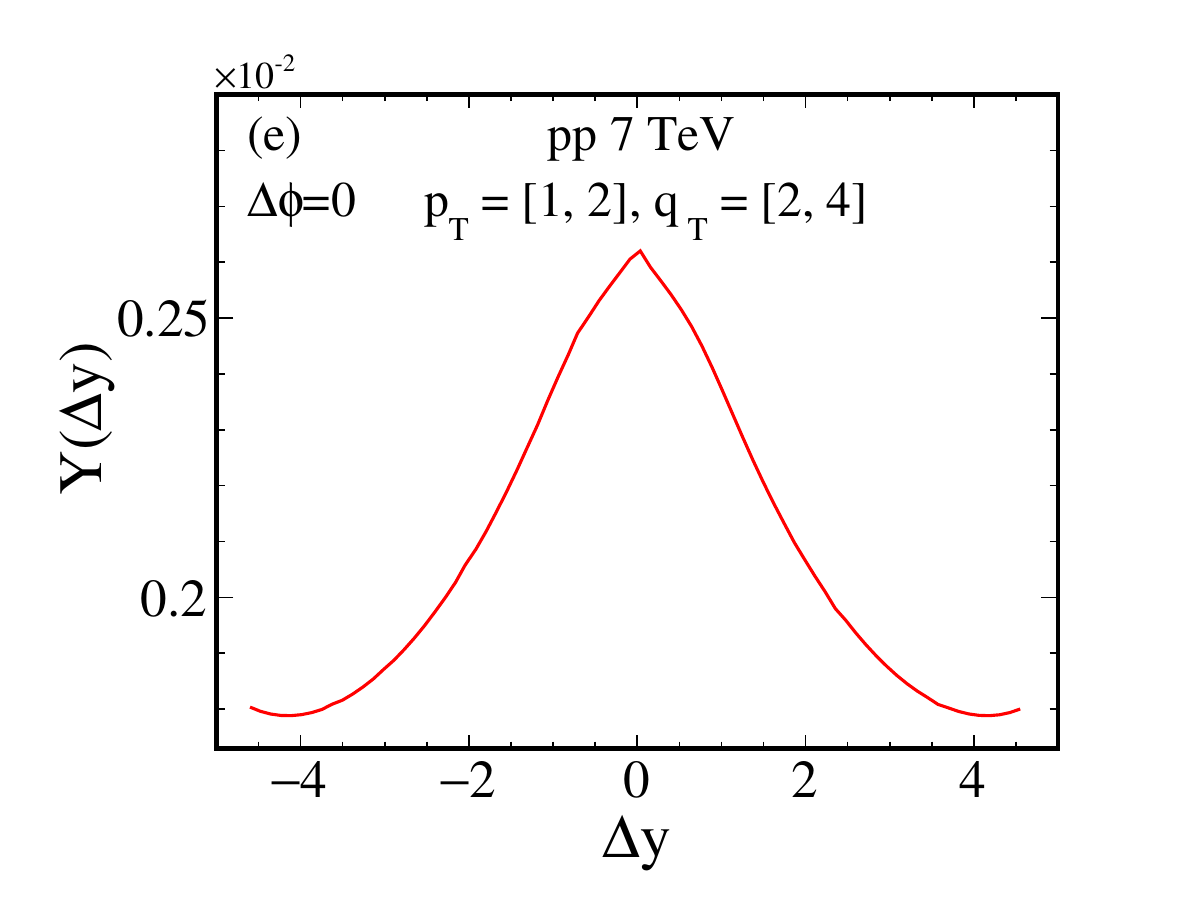}}}$ 
$\vcenter{\hbox{\includegraphics[scale=0.295]{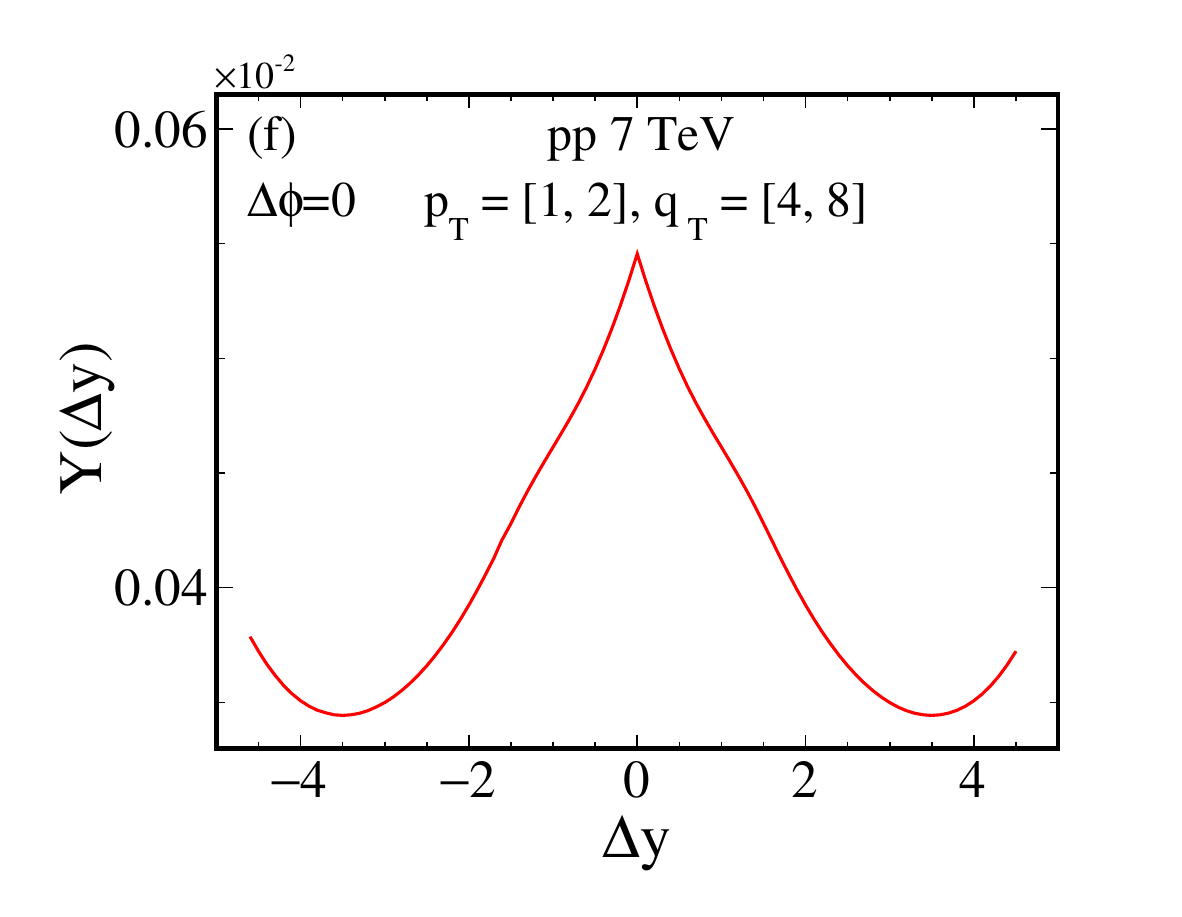}}}$
\par\end{centering}
\caption{The near-side per-trigger yield as a function of $\Delta y$ at 5.02 TeV p-Pb collisions (top row) and 7 TeV pp collisions (bottom row) within the rapidity windows $[-2.4, 2.4]$, (a) the red curve and black dotted line represent the ridge yield in transverse momentum interval $p_{\rm T}$($q_{\rm T})\in$[1, 2] and [2, 3] GeV$/c$, respectively, the ridge yield for transverse momentum interval $p_{\rm T}\in$[1, 2] GeV$/c$, $q_{\rm T}\in$[2, 4] GeV$/c$ (b), for $p_{\rm T}\in$[1, 2] GeV$/c$, $q_{\rm T}\in$[4, 8] GeV$/c$ (c), for $p_{\rm T}$($q_{\rm T})\in$[1, 2] GeV (d), for $p_{\rm T}\in$[1, 2] GeV$/c$, $q_{\rm T}\in$[2, 4] GeV$/c$ (e) and for $p_{\rm T}\in$[1, 2] GeV$/c$, $q_{\rm T}\in$[4, 8] GeV$/c$ (f), respectively.}
\end{figure*}
As Fig.~1(a) shows, at the rapidity window of $[-2.4, 2.4]$ (the CMS acceptance), the per-trigger yield in the $\Delta y$-$\Delta\phi$ plane for 5.02 TeV p-Pb collisions is given. It is similar to the results of pp collisions we obtained in the CGC framework (see Fig.~1(d) in ref.~\cite{Zhang-2}). In contrast, the ridge structure in p-Pb collisions is more prominent. In the $\Delta\phi$ direction, the per-trigger yield has two peaks of equal height at $\Delta\phi=0$ and $\pi$. The two peaks are called azimuthal collimation which is intrinsic to glasma dynamics~\cite{Dusling-2013-1,Dusling-2013-2,Dusling-2013-3}. It contributes to the well-known collectivity in small systems. 

In the $\Delta y$ direction, the ridge yield $Y$ shows first fall and then rise with $|\Delta y|$, i.e., a visible rebound after the plateau in the rapidity direction is presented. In order to observe the trend of the ridge yield $Y$ more clearly, a projection to $\Delta y$ axis is made and shown in Fig.~1(b). As the red curve shows, the near-side yield as a function of $\Delta y$ for 5.02 TeV p-Pb collisions presents a rebound at $|\Delta y|\approx3.5$ after a plateau within $2<|\Delta y|<3.5$. This trend is consistent with observations in pp collisions~\cite{CMS-2011-pp-7}, due to the absence of p-Pb data and the similarity between the ridge in pp and p-Pb collisions. We desire this rebound will be verified in future studies.

The per-trigger yield within the rapidity window of $[-0.9, 0.9]$ (the ALICE acceptance) is not shown here. Its characteristics are very similar to the results for $|\Delta y|<2$ in Fig.~1(a), without the plateau and rebound in the rapidity direction. Previous studies have indicated that glasma graphs exhibit significant short-range rapidity correlations~\cite{Zhao-2}. Consequently, the longitudinal structure of the two-dimensional distributions is less as flat as the ALICE data~\cite{ALICE-2013-pPb-5.02}. For the rapidity gap $|\Delta y|<2$, the two-dimensional distributions from the CGC framework are not directly comparable with data.

In order to see how the rebound of long-range rapidity correlations changes with colliding energy, the ridge yields $Y$ at $\sqrt{s}=5.02$ TeV and 8.16 TeV are presented in Fig.~1(b). The red curve (representing 5.02 TeV) nearly coincides with the black dotted line (representing 8.16 TeV) for $|\Delta y|<2$, but deviates significantly for $|\Delta y|>2$. The red curve (for 5.02 TeV) donates a noticeable rebound at $|\Delta y|\approx3.5$ after a plateau within $2<|\Delta y|<3.5$. The black dotted line (for 8.16 TeV) falls first then rebounds at $|\Delta y|\approx3.8$. The positions of correlation rebound at two energies are different. Figure.~1(b) indeed shows the correlation rebound shifts to larger rapidity gap at higher collision energies. This occurs because the gluon correlation is governed by the uGD. The uGD peaks at $Q_{\rm s}$ and weakens away from $Q_{\rm s}$, which is $x$ dependent~\cite{Zhao-1}. Gluons with large $x$ representing source gluons, locate in forward/backward rapidity region. Gluons with small $x$ representing radiated gluons, locate in central rapidity region~\cite{Zhang-1}. The rebound at large-rapidity gap is caused by the strong correlations between small $x$ radiated gluons and large $x$ source gluons. At fixed $x$, Eq.~(1) shows that the rapidity $y_q$ increases while $y_p$ decreases as $\sqrt{s}$ increases. The correlation rebound must move to larger rapidity gap($\Delta y=y_q-y_p$).

\section{Transverse momentum dependence of long-range rapidity correlations}

As mentioned before, the gluon correlation is governed by the uGD. The uGD peaks at $Q_{\rm s}$, which is $x$ dependent. Eq.~(1) shows that $x$ is proportional to the transverse momentum $p_{\rm T}$ at fixed $y$ and $\sqrt{s}$. The transverse momentum can also affect the correlation rebound. In the following, it is interesting to see how the rebound of large-rapidity correlations changes with the transverse momentum in symmetric pp vs. asymmetric p-Pb collisions. In order to systematically study the transverse momentum dependence, the transverse momenta of two particles are set to identical and different intervals, respectively. The results are shown in Fig.~2. The rapidity correlations within $p_{\rm T}$($q_{\rm T}$)$\in$$[1, 2]$ GeV$/c$ (the red curve in Fig.~2(a)) completely reproduce the trend of the red curve for $p_{\rm T}$($q_{\rm T}$)$\in$$[1, 3]$ GeV$/c$ in Fig.~1(b). The rebound of the red curve at $|\Delta y|\approx3.5$ in Fig.~1(b) is dominated by the transverse momentum interval $[1, 2]$ GeV$/c$. The correlations at  $p_{\rm T}$($q_{\rm T}$)$\in[2, 3]$ GeV$/c$ (the black dotted line in Fig.~2(a)) do not show any rebound trends. It indicates that the rebound of rapidity correlations at $|\Delta y|\approx3.5$ is most obvious at $p_{\rm T}$($q_{\rm T}$)$\sim Q_{\rm sA}+Q_{\rm sB}\approx 2$ GeV$/c$, where $Q_{\rm sA(B)}$ denotes the saturation momentum of the projectile (A) or target (B). As we know, $Q_{\rm s}$ is $\sqrt{s}$ dependent. At 5.02 TeV, the saturation momentum of proton and lead nucleus are $Q_{\rm sp}\approx 0.8$ GeV$/c$ and $Q_{\rm sA}\approx 1.2$ GeV$/c$ (see Fig.~2(a) in ref.~\cite{Zhao-1}). This is consistent with the existing experimental result that ridge yield gets the maximum within $[1, 2]$ GeV$/c$ of particle transverse momentum~\cite{CMS-2010-pp}. For this reason, we always choose $p_{\rm T}\in$[1, 2] GeV$/c$ in the following. 

As Fig.~2(b) shows, the ridge yield $Y$ at $q_{\rm T}\in$[2, 4] GeV$/c$ falls first then rises with $|\Delta y|$, showing a visible rebound after the plateau in the rapidity direction. A striking difference from Fig.~2(a) is that the rapidity correlations become asymmetric. The correlation rebound at $\Delta y \approx-3.5$ is higher and more obvious than at $\Delta y \approx3.5$, indicating a stronger correlation at $\Delta y \approx-3.5$. It is further observed that the distribution of $Y(\Delta y)$ reverses when the $p_{\rm T}$ and $q_{\rm T}$ intervals are swapped, i.e., $p_{\rm T}\in$[2, 4] GeV$/c$ and $q_{\rm T}\in$[1, 2] GeV$/c$. In this case, the correlation rebound at $\Delta y \approx-3.5$ is lower and less pronounced than $\Delta y \approx3.5$. A nontrivial $p_{\rm T}$-dependence is observed in asymmetric rapidity correlations. 

The rapidity correlations at $q_{\rm T}\in$[4, 8] GeV$/c$ are illustrated in Fig.~2(c). The curve in Fig.~2(c) is similar to the curve in Fig.~2(b), with the exception of the amplitude. As Fig.~2(c) shows, a rebound occurs at $|\Delta y|\approx3.5$ after the plateau within $2<|\Delta y|<3.5$. The correlation rebound at $\Delta y \approx-3.5$ is stronger and more distinct than at $\Delta y \approx3.5$. 

Comparing Figs.~2(a), 2(b) and 2(c), it is observed that the amplitude of $Y(\Delta y)$ decreases as $q_{\rm T}$ increases. The rapidity correlations are asymmetric in p-Pb collisions at different $p_{\rm T}$ and $q_{\rm T}$ intervals, but become symmetric when the intervals of $p_{\rm T}$ and $q_{\rm T}$ overlap. Large-rapidity ridge correlations exhibit a strong $p_{\rm T}$-dependent asymmetry. 

To verify whether the $p_{\rm T}$-dependent asymmetry are unique to p-Pb collisions in the CGC framework, the rapidity correlations for pp collisions at 7 TeV within rapidity window $[-2.4, 2.4]$ are calculated for comparison. The rapidity correlations for pp collisions with $p_{\rm T}$($q_{\rm T})\in$[1, 2] GeV$/c$ are shown in Fig.~2(d). For  $q_{\rm T}\in$[2, 4] GeV$/c$ and [4, 8] GeV$/c$ ($p_{\rm T}\in$[1, 2] GeV$/c$), the correlations are illustrated in Figs.~2(e), 2(f), respectively. As Figs.~2(d), 2(e) and 2(f) show, all three curves have similar structures. They first fall then rise with $|\Delta y|$, showing a rebound after the plateau in the rapidity direction. Additionally, the amplitude of $Y(\Delta y)$ decreases as $q_{\rm T}$ increases. In contrast to p-Pb collisions, the rapidity correlations in pp collisions are always symmetric, regardless of the variations in the transverse momenta of two particles. A unique finding in Figs.~2 is that the rapidity correlations demonstrate strong $p_{\rm T}$ dependence in p-Pb collisions, but are $p_{\rm T}$-independent in pp collisions.

The transverse momentum dependence of large-rapidity ridge correlations can be well explained within the CGC framework. The per-trigger yield is proportional to the correlated two-gluon inclusive distributions, i.e., Eq.~(7), which can be expressed by convolutions of four uGDs~(Eq.~(9)-Eq.~(13)), e.g.,
\begin{equation}
\Phi^2_{\rm A}(y_{\rm p},\boldsymbol{k}_{\rm T})\Phi_{\rm B}(y_{\rm p},\boldsymbol{p}_{\rm T}-\boldsymbol{k}_{\rm T})\Phi_{\rm B}(y_{\rm q},\boldsymbol{q}_{\rm T}-\boldsymbol{k}_{\rm T}).
\end{equation}
The uGD ($\Phi$) depends on $x$ and peaks at $Q_{\rm s}$, transverse momentum far from $Q_{\rm s}$ contributes little to the correlation. The strongest correlation~\cite{Dumitru-2011,Zhao-1,ZhangHY} requires simultaneously, 
\begin{equation}
|\bo{k}_{\rm T}|\sim Q_{\mathrm{sp}},\quad |\bo{p}_{\rm T}-\bo{k}_{\rm T}|\sim Q_{\mathrm{sA}}, \quad |\bo{q}_{\rm T}-\bo{k}_{\rm T}|\sim Q_{\mathrm{sA}.}
\end{equation}
In the context, the maximum of the correlation should be near $|\bo{p}_{\rm T}|\sim|\bo{q}_{\rm T}|\sim Q_{\rm sp}+Q_{\rm sA}\sim 2$ GeV$/c$ in pA collisions. As $q_{\rm T}$ increases and moves farther from $Q_{\rm s}$, the correlation becomes weaker. Therefore, the amplitude of $Y(\Delta y)$ in Figs.~2 decreases with $q_{\rm T}$ increases. 
\begin{table*}
\caption{\label{tab:ex}An example of the Bjorken $x$ values at $|\Delta y|=3.5$ in symmetric and asymmetric rapidity windows.}
\begin{ruledtabular}
\begin{tabular}{l|cc||cc}
Acceptance & \multicolumn{2}{c||}{$Y_{\rm w}=[-2.4,2.4]$} & \multicolumn{2}{c}{$Y_{\rm w}=[-2.865, 1.935]$}\\ \hline
Rapidity gap & $\Delta y=-3.5$ & $\Delta y = 3.5$ & $\Delta y=-3.5$ & $\Delta y= 3.5$\\ \hline
\multirow{2}{4em}{$x$-values} & $x_q=\frac{q_T}{\sqrt{s}}e^{2.0}$ & $x_q=\frac{q_T}{\sqrt{s}}e^{2.0}$ & $x_q=\frac{q_T}{\sqrt{s}}e^{2.4}$ & $x_q=\frac{q_T}{\sqrt{s}}e^{1.0}$\\
 & $x_p = \frac{p_T}{\sqrt{s}}e^{1.5}$ & $x_p = \frac{p_T}{\sqrt{s}}e^{1.5}$ & $x_p = \frac{p_T}{\sqrt{s}}e^{1.0}$ & $x_p = \frac{p_T}{\sqrt{s}}e^{2.4}$ \\
\end{tabular}
\end{ruledtabular}
\end{table*}

Let's clarify the reasons for the formation of asymmetric rapidity correlations. As mentioned earlier, the gluon correlation is determined by the uGD. The uGD depends on $x$. In a symmetric rapidity window, like the CMS acceptance $[-2.4, 2.4]$, the rapidity of two particles can take opposite values at same $|\Delta y|$, e.g., $y_{q}=-2(2)$ and $y_{p}=1.5(-1.5)$ for $\Delta y$=-3.5(3.5). Table.~1 shows that $x_q$, $x_p$ at $|\Delta y|$=3.5 respectively correspond to equality. It means that the correlations are always identical at same $|\Delta y|$. The rapidity correlations are thus symmetric in symmetric pp and AA collisions systems. 

In asymmetric collision systems, like pA collisions, the asymmetry of the colliding nucleus leads to different rapidity windows in the laboratory and center-of-mass frames. For instance, the CMS acceptance is $[-2.4, 2.4]$ in laboratory frame, but is $[-2.865, 1.935]$ in center-of-mass frame. Theoretical calculations carry out in the center-of-mass frame, so the rapidity window is $[-2.865, 1.935]$ in our calculation. At large-rapidity gap $|\Delta y|$, the rapidity of two particles can only swap magnitudes, not signs, e.g., $y_{q}=-2.4(1.0)$ and $y_{p}=1.0(-2.4)$ for $\Delta y=-3.5(3.5)$. As shown in Table.~1, although $x_q, x_p$ are different at $|\Delta y|$=3.5, $x_q(x_p)$ at $\Delta y=-3.5$ always equals to $x_p(x_q)$ at $\Delta y=3.5$ when $p_{\rm T}$ and $q_{\rm T}$ are identical. The correlation rebound is governed by the convolutions of four uGDs which depends on $x$-component of two selected gluon like Eq.~(22). Therefore, the rapidity correlations are identical at same $|\Delta y|$ and are symmetric shown in Fig.~2(a). 
 
When $p_{\rm T}$ and $q_{\rm T}$ are unequal, the value of $x_q$ and $x_p$ at $\Delta y$=3.5 are distinct from those at $\Delta y$=-3.5, and so does correlation strength. We can find $x_q=0.0022$ and $x_p=0.0047$ at $\Delta y=3.5$, $x_q=0.0094$ and $x_p=0.0011$ at $\Delta y=-3.5$ for $p_{\rm T}=2$ and $q_{\rm T}=4$. From a physical perspective, $x_q$ is closer the large $x$($x>0.01$) at $\Delta y=-3.5$, leading to the formation of a quasi large-$x$ and small-$x$ correlation pair. Consequently, the correlations at $\Delta y =-3.5$ are stronger than those at $\Delta y=3.5$. This naturally explains the observed asymmetry. Moreover, $y_q$ decreases with $\Delta y$ at $\Delta y<-3.5$, $x_q$ increases with $\Delta y$ decreases and is closer the large $x$($x>0.01$). As a result, the correlation rebound becomes more significant at $\Delta y <-3.5$, as shown in Fig.~2(b) and 2(c). When the intervals of $p_{\rm T}$ and $q_{\rm T}$ are swapped, i.e., $p_{\rm T}\in[2, 4]$ and $q_{\rm T}\in[1, 2]$, then a quasi large-$x$ and small-$x$ correlation pair is formed at $\Delta y=3.5$ instead of at $\Delta y=-3.5$. This causes the distribution of $Y(\Delta y)$ to reverse. In this manuscript, we calculated that the lead nucleus along the beam direction collide with proton, swapping the proton and lead nucleus will reverse the correlation patterns. 

The asymmetry in rapidity correlations originates from the inherent asymmetry of the colliding hadron or nucleus in rapidity. The asymmetric rapidity window in the center-of-mass frame causes an asymmetric selections for particle pairs with the same $|\Delta y|$, resulting in different $x$ values and correlation strength at same $|\Delta y|$. This ultimately creates the asymmetric rapidity correlations.  

\section{summary and discussions}
Within the framework of CGC, we study the long-range rapidity correlations in p-Pb collisions at $\sqrt{s_{\mathrm{NN}}}=5.02$ TeV by using the exact normalization scheme proposed in our previous paper~\cite{Zhang-2}. After this normalization, the ridge correlation rebounds after bottoming. A clear rebound structure appears at large-rapidity region. The correlation rebound is found to appear around the sum of the saturation momentum of the projectile and target. In addition, the rebound moves to larger rapidities at higher colliding energies. These predictions await for experimental verification. 

It is further found that the near-side rapidity correlations demonstrate the $p_{\rm T}$-dependent asymmetry in p-Pb collisions. The formation of asymmetric rapidity correlations stems from the kinematic dependence of the uGD, which peaks at the saturation momentum $Q_s(x)$. Eq.~(1) shows rapidity as well as transverse momentum can affect the Bjorken $x$. The asymmetric rapidity window and the different transverse momenta of two particles jointly result in the asymmetric rapidity correlations. Physically, gluons with different $x$ degree of freedom reflect different stages of partonic evolution and correspond to different rapidity regions. The correlation rebound essentially probes strong correlations between large $x$ source gluons and their small $x$ radiated descendants. 

The physics of correlations of gluons at different rapidity regions essentially reflect correlations of gluons of different $x$. The correlation patterns are dependent on the $x$-component of two selected gluons. This physical picture may also understand the ridge correlations of two particles at different rapidity regions in pA collisions given by ALICE collaborations~\cite{Alice-Ultra}. 

The rebound structure at large rapidity separation and the asymmetric ridge correlation are caused by the stronger correlation between source gluons and radiated gluons. These features are independent of the speed of the small $x$ evolution and the selection of initial conditions. Therefore, the selection of running coupling kernels and initial conditions does not influence the qualitative feature of the results. 

So far the correlation patterns calculated are for gluons. The fragmentation functions reflecting final hadronization should also be considered. As ref.~\cite{Dusling-2012} shows, including fragmentation functions brings integrations over $z$, the transverse momentum fraction of the produced hadron with respect to that of the fragmenting gluon, the rapidity correlations of two hadrons at a certain transverse momentum, e.g., $p_0$, take into account the contributions of all gluons with transverse momenta larger than $p_0$. Therefore fragmentation functions only have a major impact on transverse momentum dependence. The correlation patterns are dependent on the $x$-component of two selected gluon. According to , the influence of transverse momentum on Bjorken $x$ is much smaller than that of rapidity. This should not change qualitatively the rebound features at parton level. From the results, the rebound structures are still prominent and rapidity correlations $Y(\Delta y)$ are always asymmetric even the increase of $q_{\rm T}$ in Figs.~2. Therefore the correlation patterns as a function of rapidities would remain in the final state. Besides, the nonperturbative effects specifying the correction to the $k_{T}$ factorized uGD description is regarded as constant~\cite{Mace-2019}. 

In this manuscript, only glasma graph is calculated which results in both the near-side and the away-side ridge, i.e., the so-called double ridges. Other effects also give significant contributions to the two-particle correlations, e.g., Mueller-Navelet di-jet and jet shower effects. It is known that di-jet mainly contributes to the away-side ridge and jet shower contributes to short-range correlations of the near side. When long-range rapidity correlations of the near side are considered, both effects are negligible. Further observation the rebound structure of large-rapidity ridge correlations in experiments is a direct test of the CGC mechanism vs. the final-state effects, and is thus very meaningful and interesting.

\section*{Acknowledgement}
This research was partially supported by the National Key Research and Development Program of China, grant number 2022YFA1604900; the National Natural Science Foundation of China, grant number 12275102; the Fundamental Research Funds of China West Normal University, grant number 22kE042; the Scientific Research and Innovation Team Program of Sichuan University of Science and Engineering(No.SUSE652A001). The numerical simulations have been performed on the GPU cluster in the Nuclear Science Computing Center at Central China Normal University (NSC3).

\providecommand{\href}[2]{#2}\begingroup\raggedright\endgroup
\end{document}